\documentstyle[12pt]{article}
\begin{document}
\baselineskip 6mm

\hfill~{KLTE-DTP/1998/1}

\begin{center}
\Large{{\bf
 Generalized Rarita-Schwinger Equations}}\\
\end{center}
\begin{center}
I. Lovas       \\
 Research Group for Theoretical Physics (MTA-KLTE)
 \\  Kossuth  Lajos University,
       Debrecen, Hungary\\
 
\begin{center}
K. Sailer    \\
\end{center}
 Department of Theoretical Physics,
 \\  Kossuth  Lajos University,
       Debrecen, Hungary\\
\end{center}
\begin{center}
W. Greiner \\
 Institute of Theoretical Physics
 \\  Johann Wolfgang Goethe  University,
     Frankfurt am Main, Germany\\
\end{center}
\vspace{0.5cm}
\vspace{0.5cm}
To be published in Heavy Ion Physics.\\
\vspace{0.5cm}
\vspace{0.5cm}
\\
{\bf Abstract}
The Rarita-Schwinger equations are  generalised for the
delta baryon having spin 3/2 and isospin 3/2.
 The coupling of the nucleon and the delta fields is studied.
A possible generalisation of the Walecka model is proposed.

\section{Introduction}
\indent

     In a previous paper
\cite{Lov97} the
problem of spin $3/2$ has been discussed. More specificaly
the generalisation of the
Rarita-Schwinger equations has been presented for  particles
having isospin as internal degree of freedom.
 
The description
 of particles with
 spin 3/2 can be given  by the help of the Bargmann-Wigner
\cite{Bar41} procedure introducing the  multispinor fields
\begin{equation}
\Psi_{\alpha \beta \gamma}(x),
\end{equation}
which must be completely symmetric in all of their  indices and
the Dirac-equation must be satisfied for all
indices:
\begin{eqnarray}
 (\gamma_\mu \partial^\mu + M)_{\alpha \alpha'}  \Psi_{\alpha'
\beta \gamma
}(x)=0, \\
 (\gamma_\mu \partial^\mu + M)_{\beta \beta'}
\Psi_{\alpha \beta' \gamma
}(x)=0, \\
 (\gamma_\mu \partial^\mu + M)_{\gamma \gamma'}  \Psi_{\alpha
\beta \gamma'
}(x)=0.
\end{eqnarray}
These requirements lead to the Rarita-Schwinger equations given
\cite{Rar41}
by

\begin{equation}
 (  \gamma_\kappa \partial^\kappa +M)
\psi^\mu(x)=0 .
\end{equation}
 
\begin{equation}
\label{cons}
   \gamma_\mu
 \psi^\mu
 =0.
\end{equation}
The completely symmetric function
$\Psi_{\alpha \beta \gamma}(x)$
can be built from
$\psi^\mu(x)$.
As it is seen all of the vector components of $\psi^\mu(x)$
satisfy
 Dirac-type equations, however they are not independent, they
must fulfill the  constraint given by Eq. (\ref{cons}).
This constraint
coming from the  symmetry condition projects out the spin 1/2
components and the remaining components describe particles with
spin 3/2.
 
Here we use the notation  $x_4=it$ and summation is implied for
every two identical indices.
The Dirac-matrices $\gamma_\mu$ satisfy the
anticommutator relations:
 
\begin{equation}
\gamma_\mu \gamma_\nu+\gamma_\nu
\gamma_\mu=2 \delta_{\mu \nu} \quad
(\mu,\nu=1,2,3,4)
\end{equation}
 
\section{Rarita-Schwinger Equations for Particles with Isospin}
 
\subsection{Symmetry Conditions}
\indent
 
Using the Bargmann-Wigner procedure
\cite{Bar41}
we have generalised the Rarita-Schwinger equations
assuming  that all the
indices $\alpha$, $\beta$, $\gamma$ of
the field
$\Psi_{\alpha \beta \gamma}(x)$
 contain an isospin index too, labeling the eigenstates of the
third component of isospin 1/2.
 The field
$\Psi_{\alpha \beta \gamma}(x)$
must be
completely  symmetric in its indices.
In the 8-dimensional space there exist 36 linearly independent
 matrices which are
 symmetric in $\alpha$ and $\beta$ and can be
given as follows:
\begin{equation}
(T C)_{\alpha \beta}, \qquad
(i \gamma_5 T C)_{\alpha \beta} , \qquad
( \gamma_\mu i \gamma_5 T C)_{\alpha \beta},\]
\[( \gamma_{\mu}\tau_m T C)_{\alpha \beta}, \qquad
( {\mbox{\small$\frac12$}} \Sigma_{\mu \nu} \tau_m T C)_{\alpha
\beta}.
\end{equation}
Here  the  matrices
$ \Sigma_{\mu \nu}$,
$C$ and  $T$
  are  defined by
\begin{equation}
\Sigma_{\mu \nu}=(\gamma_\mu \gamma_\nu-\gamma_\nu
\gamma_\mu)/(2 i),
\end{equation}
\begin{equation}
 C=\gamma_2 \gamma_4,
\end{equation}
\begin{equation}
T=\tau_1   \tau_3,
\end{equation}
with  the  Pauli-matrices
 $\tau_m$  $(m=1,2,3)$.

Using these
matrices the multispinor field is defined in the following form:
\begin{equation}
  \Psi_{\alpha \beta \gamma}(x)=
 (T C)_{\alpha \beta} \psi^{0 }_{\gamma} +
(i \gamma_5 T C)_{\alpha \beta} \psi^{5}_{\gamma}+
( \gamma_\mu i \gamma_5 T C)_{\alpha \beta} \psi^{\mu
5}_{\gamma}  \]
\[  + ( \gamma_{\mu}\tau_m T C)_{\alpha \beta} \psi^{\mu m}_{\gamma}+
( {\mbox{\small$\frac12$}}\Sigma_{\mu \nu}\tau_m T C )_{\alpha
\beta} \psi^{\mu
\nu m}_{\gamma}.
\end{equation}
  The task is to determine the functions
$\psi^{0}(x)$,
$\psi^{5}(x)$,
$\psi^{\mu 5}(x)$,
$\psi^{\mu m}(x)$
and
$\psi^{ \mu\nu m}(x).$
Up till now
the field
$ \Psi_{\alpha \beta \gamma}(x)$
is symmetric only in the
indices $\alpha$ and $\beta$.
To guarantee the complete  symmetry one must eliminate the
 components antisymmetric in
respect of the second and third indices. This
requirement is
fulfilled  if the  contraction of $ \Psi_{\alpha \beta \gamma}(x)$ with
 all of the matrices
antisymmetric in the indices
$\beta$ and $ \gamma$ are vanishing.
In 8-dimensional  space there exist
28 antisymmetric, linearly
independent matrices, which can be defined as follows:
\[  (C^{-1} T^{-1} \tau_n )_{\beta \gamma}, \qquad
(C^{-1} T^{-1} i \gamma_5 \tau_n)_{\beta \gamma}, \qquad
(C^{-1} T^{-1}  \gamma_{\lambda} i \gamma_5 \tau_n)_{\beta
\gamma},\]
\begin{equation}
 (C^{-1} T^{-1} \gamma_{\lambda})_{ \beta \gamma}, \qquad
(C^{-1} T^{-1} {\mbox{\small$\frac12$}}\Sigma_{\lambda \rho}
)_{\beta \gamma}.
\end{equation}
Thus the  requirement of the complete symmetry leads to the
following 28  conditions:
 
\begin{eqnarray}
\gamma_{\mu} \psi^{\mu m} &=&
\tau^m (\psi^{0 } -
i \gamma_5  \psi^{5}),
\\
 {\mbox{\small$\frac12$
}}\Sigma_{\mu \nu}
\psi^{\mu \nu m} &=&
\tau^m (\psi^{0 } +
i \gamma_5  \psi^{5}),
\\
i\gamma_\mu
\psi^{\mu \nu m}&=&
-\psi^{\nu m} +
\tau^m( \tau_n \psi^{\nu n}+
\gamma^\nu i\gamma_5 \psi^{5}),
\\
i\gamma_5
\psi^{\mu 5}&=&
-\tau_m \psi^{\mu m} +
\gamma^\mu(  \psi^{0}-
i\gamma_5 \psi^{5}),
\\
\tau_m
\psi^{\mu \nu m}&=&
i\tau_m(
\gamma^\nu  \psi^{\mu m}-
\gamma^\mu   \psi^{\nu m}) +
\Sigma^{\mu \nu} (-\psi^0+2i\gamma_5 \psi^5).
\end{eqnarray}

\subsection{Decomposition of the Dirac Equation}
\indent
 
      As it was stated in the Introduction the field
$\Psi_{\alpha \beta \gamma}(x)$
must satisfy the Dirac-equation with respect of its all three
indices.
 
  By  applying the Dirac-operator
$(\gamma_\kappa \partial^\kappa +M)$ on the first index
$\alpha$ of
$\Psi_{\alpha \beta \gamma}(x)$
the following equation is obtained:

\begin{equation}
(\gamma_\kappa \partial^\kappa +M)_{\alpha \alpha'}
 (T C)_{\alpha' \beta}
\psi^{0 }_{\gamma}(x) +\]
 
\[(\gamma_\kappa \partial^\kappa +M)_{\alpha \alpha'}
(i \gamma_5 T C)_{\alpha' \beta}
\psi^{5}_{\gamma}(x)+\]
 
\[(\gamma_\kappa \partial^\kappa +M)_{\alpha \alpha'}
 ( \gamma_\mu i \gamma_5 T C)_{\alpha' \beta}
\psi^{\mu 5}_{\gamma}(x)+\]
 
\[(\gamma_\kappa \partial^\kappa +M)_{\alpha \alpha'}
( \gamma_{\mu}\tau_m T C)_{\alpha' \beta}
\psi^{\mu m}_{\gamma}(x)+ \]
 
\[(\gamma_\kappa \partial^\kappa +M)_{\alpha \alpha'}
(  {\mbox{\small$\frac12$}}\Sigma_{\mu \nu}\tau_m T C )_{\alpha'
\beta}
\psi^{\mu \nu  m}_{\gamma}(x)
=0.
\end{equation}
 
This equation is fulfilled if its 36 linearly
independent components vanish. To separate  these
components   we
contract  this equation with the following 36
matrices

\begin{equation}
(C^{-1} T^{-1}
)_{\beta \alpha },\qquad
(C^{-1} T^{-1}
i \gamma_5 )_{\beta \alpha } , \qquad
(C^{-1} T^{-1}
 \gamma_\lambda i \gamma_5 )_{\beta \alpha },\]
 
\[(C^{-1} T^{-1}
 \gamma_{\lambda}\tau_n)_{\beta \alpha }, \qquad
(C^{-1} T^{-1}
 {\mbox{\small$\frac12$}} \Sigma_{\lambda \rho}\tau_n )_{\beta \alpha
},
\end{equation}
which are linearly independent and symmetric in their indices
$\beta$ and $\alpha$.
 
The result is the following 36 equations:
\begin{eqnarray}
M\psi^0&=&0,
\\
M\psi^5&=&-\partial_\kappa \psi^{\kappa 5},
\\
M\psi^{\lambda 5}&=&-\partial^\lambda \psi^{5},
\\
M\psi^{\lambda n}&=&i \partial_\kappa \psi^{\kappa \lambda n},
\\
M\psi^{\rho \lambda n}&=&i(\partial^\lambda \psi^{\rho n}-\partial^\rho
\psi^{\lambda n}).
\end{eqnarray}

As it is seen
the resulting equations form a coupled set of differential
equations for the
determination of the functions
$\psi^{0}(x)$,
$\psi^{5}(x)$,
$\psi^{\lambda 5}(x)$,
$\psi^{ \lambda n}(x)$ and
$\psi^{\rho \lambda n}(x).$
 
The functions $\psi^{\rho\lambda
n}$ are not independent quantities since they can  be derived from
$\psi^{\rho n}$ by the help of the equation (25).
The equation (21) shows that the function $\psi^{0}$ must be zero:
\begin{eqnarray}
\psi^{0}=0.
\end{eqnarray}
 
The rest of these equations together with the conditions (14)-(18)
 can
be reduced to the following set of equations:
 
\begin{eqnarray}
 (  \gamma_\kappa \partial^\kappa +M)
\psi^{\mu m}(x)&=&0 ,
\\
 (  \gamma_\kappa \partial^\kappa +M)
\psi^{\mu 5}(x)&=&0 ,
\\
 (  \gamma_\kappa \partial^\kappa +M)
\psi^{5}(x)&=&0 ,
\\
\gamma_\mu
 \psi^{\mu m}
&=& -\tau^m i\gamma_5 \psi^5,
\\
 \tau_m
 \psi^{\mu m}
&=&i\gamma_5 \psi^{\mu 5}-\gamma^{\mu}i\gamma_5 \psi^5.
\end{eqnarray}
 
   This can be simplified by
assuming trivial solutions for
$\psi^5$ and $\psi^{\lambda 5}$:
\begin{equation}
\psi^{5}(x)=0,
\qquad
\psi^{\lambda 5}(x)=0.
\end{equation}
 
  In this way the number of components is decreased,
however, the number of  independent components does not. Thus
the conditions given above and the remaining differential
equations
 are reduced
to the following set of equations:
 
\begin{eqnarray}
 (  \gamma_\kappa \partial^\kappa +M)
\psi^{\mu m}(x)&=&0 ,
\\
\gamma_\mu
 \psi^{\mu m}
&=&0,
\\
 \tau_m
 \psi^{\mu m}
&=&0.
\end{eqnarray}
 
These are  the generalised
Rarita-Schwinger equations.
 
The field
 $\psi^{\mu m}_\gamma(x)$
 is transformed  as a direct product of a Lorentz-vector
$(\mu)$
and a Dirac-spinor $(\gamma)$ multiplied with a direct product of
an isovector $(m)$ and an isospinor $(\gamma)$. As a consequence
of
these properties it contains components of spin (3/2 and 1/2) and
isospin (3/2 and 1/2).
The constraint (34) projects out from $\psi^{\mu m}$ the spin 1/2
components, while the constraint (35) projects out the isospin 1/2
components.
 
\section{Solutions of the Generalised Rarita-Schwinger Equations}

In this Section we study the plane wave type solutions of the
generalised Rarita-Schwinger equations, defined by
\begin{eqnarray}
\psi^{\mu m} (x) = u^{\mu m} (\vec k,E) e^{ikx}
\end{eqnarray}
The amplitude $u^{\mu m} (\vec k, E)$ must satisfy the following algebraic
equation:
\begin{eqnarray}
(i \vec k \vec \gamma - E \gamma_{4} + M) u^{\mu m} (\vec k, E ) = 0
\end{eqnarray}
If $ E>0$ then in the rest frame, where $\vec k = 0$  and $ E= M$
this equation is reduced to the following condition:
\begin{eqnarray}
\gamma_{4} u^{\mu m}(0) = u^{\mu m}(0)
\end{eqnarray}
where the shorthand notation
$u^{\mu m} (\vec k=0, E=M) = u^{\mu m}(0)$
has been introduced.
 This condition  means that the small components  vanish.
 The constraints
 
\begin{eqnarray}
\gamma_{\mu} u^{\mu m}(0) = 0 \; \; \; \mbox{and} \; \; \;
\tau_{m} u^{\mu m}(0)
=
0
\end{eqnarray}
lead to the following results
 
\begin{eqnarray}
u^{4 m}(0) = 0, \;\;
\sigma_{a} u^{a m}(0) = 0, \; \;
\tau_{m} u^{a m}(0) = 0,
\end{eqnarray}
where              $a =1,2,3$ and $m =1,2,3$.
 
    Apparently we have $9 \times 4 = 36$ constraints for
the $12 \times 4 = 48$
components of
$u^{\mu m}(0)$.
      However these constraints are not independent:
         $ \sigma_{a} \tau_{m} u^{am}(0)
=\tau_{m} \sigma_{a} u^{am}(0) $,
consequently the number of the constraints is only 32 and so the
number of the  independent components of
$u^{\mu m}$ in
the rest frame is 16. This means that the Bargmann--Wigner procedure
provides  the description of a particle with spin 3/2 and isospin 3/2,
as it will be discussed below.
 
   First of all   we note that it is possible to
introduce total isospin eigenstates instead of
 $u^{\mu m} (\vec k, E)$  by the help of Clebsch-Gordan coefficients. If
the total isospin is 1/2 then the eigenstate vanishes identically
because of the constraint (35). The other linearly independent
eigenstate characterised by the total isospin 3/2 can be given as
\begin{equation}
u^{\mu t}(\vec k,E)=\sum (1 m \frac{1}{2} m_{\gamma}|\frac{3}{2} t)
u^{\mu m}_{m_{\gamma}}(\vec k,E),
\end{equation}
where
$m_\gamma$ is the isospinor index
suppressed up till now and the eigenvalue of the third component of the
total isospin is denoted by $t$.
 
    Similarly
to the total isospin we may introduce the total spin
eigenstates
$u^{\mu t}_{\sigma}$, where the eigenvalue of the third component of the
total spin is denoted by $\sigma$.
         For fixed  values  of $\vec k , E$ and that of the
indices $a$ and $t$
4 independent solutions, characterised by $\sigma= +3/2, +1/2, -1/2,
-3/2$ exist which can be given as follows:
\begin{eqnarray}
u_{+3/2}^{at} (\vec k, E) = \varepsilon_{+1}^{a} u^{t}_{+1/2},\\
u_{+3/2}^{4t}(\vec k, E) = 0;
\end{eqnarray}
\begin{eqnarray}
u_{+1/2}^{at} (\vec k, E) = \sqrt\frac{1}{3} \varepsilon_{+1}^{a}
u^{t}_{-1/2}
 - \sqrt\frac{2}{3} \frac{E}{M} \varepsilon_{0}^{a}
u_{+1/2}^{t} , \\
u_{+1/2}^{4t} (\vec k, E) = -i \sqrt\frac{2}{3}
\frac{|\vec k|}{M}  u^{t}_{+1/2};
\end{eqnarray}
\begin{eqnarray}
u_{-1/2}^{at} (\vec k, E) = \sqrt\frac{1}{3} \varepsilon_{-1}^{a}
u^{t}_{+1/2}
 + \sqrt\frac{2}{3} \frac{E}{M} \varepsilon_{0}^{a}
u_{-1/2}^{t},    \\
u_{-1/2}^{4t} (\vec k, E) = +i \sqrt\frac{2}{3}
\frac{|\vec k|}{M}  u^{t}_{-1/2};
\end{eqnarray}
\begin{eqnarray}
u_{-3/2}^{at} (\vec k, E) =  \varepsilon_{-1}^{a}
u^{t}_{-1/2},      \\
u_{-3/2}^{4t} (\vec k, E) = 0.
\end{eqnarray}
 
    Here the amplitudes $u_{\Sigma}^{t}$ and
$\varepsilon_{S}^{a}$ are defined as orthonormalised helicity eigenstates
obtained from the
following equations
\begin{eqnarray}
\frac{1}{2} \frac{\vec \Sigma \vec k}{|\vec k|} u^{t}_{\Sigma}
= \Sigma u^{t}_{\Sigma}, \;\;
(\Sigma = +1/2,-1/2).
\end{eqnarray}
 
\begin{eqnarray}
\frac{\vec S \vec k}{|\vec k|} \varepsilon_{S}^a  = S\varepsilon_{S}^a,\;\;
(S = -1, 0, +1).
\end{eqnarray}
 
The operators   $ \Sigma_i$ and $ S_i $   are defined as follows

\begin{eqnarray}
\Sigma_i=-i/2 \epsilon_{ijk4} \gamma_j \gamma_k,\\
(S_i)_{jk}= -i \epsilon_{ijk},
\end{eqnarray}
where the completely antisymmetric Levi-Civita symbol is denoted
by $\epsilon$.

\section{Non-Strange Baryons}
\indent

        Up till now no elementary particle with spin 3/2 has been
observed. This means that no real application exists for the
Rarita-Schwinger equations.
On the other hand among the baryons the delta resonances,
observed by Fermi \cite{Fer52}, have spin 3/2. They are, however composite
particles, built up from quarks. Nevertheless they are generally
considered
in nuclear physics as structureless objects with spin
3/2
and isospin 3/2. In the quark model of Gell-Mann
 \cite{Gel64} the delta, as a baryon, is a  bound state of three
quarks described by
the symmetric wave function
 
\begin{equation}
  [\phi_\alpha(1) \phi_\beta(2) \phi_\gamma(3)]
\end{equation}
where the single particle functions
   $\phi_\alpha(i)$ depend on the space, spin and isospin
variables of a single quark.
 It has to be noted that the antisymmetry of the three-quark
state is guaranteed by the antisymmetric color factor which is
omitted here. The quarks are confined to the
volume
of the delta  and in the ground state they are in a state symmetric
with respect of the interchange of space coordinates,
 consequently the ground state must
be  symmetric in the spin-isospin variables.
Among the circumstances of low energy phenomena, when the quark
degrees of freedom are frosen yet, we can forget about the space-time
coordinates of the
individual quarks and can introduce the space-time coordinates
 $x$ of the delta as a whole. Then the delta can be
described by the wave function
$ \Psi_{\alpha\beta\gamma}(x)$ which is  symmetric
 with respect of the interchange of the spin-isospin indices
$\alpha, \beta$ and $\gamma$.
   The nucleon in its ground state having spin 1/2 and isospin 1/2 can
be described similarily by a function $\Psi_{\alpha}$   satisfying
the Dirac equation.
 
The relativistic wave functions of the non-interacting, non-strange
baryons,
 namely those of  the nucleon and the
delta
satisfy the following Dirac-type equations:
 
\begin{eqnarray}
 (  \gamma_\kappa \partial^\kappa +M_N)
\psi(x)&=&0 ,\\
 (  \gamma_\kappa \partial^\kappa +M_{\Delta})
\psi^{\mu m}(x)&=&0 ,
\\
\gamma_\mu
 \psi^{\mu m}
&=&0,
\\
 \tau_m
 \psi^{\mu m}
&=&0.
\end{eqnarray}

\section{Quantumhadrodynamics}
\indent
 
   In order to have a complete quantum field theory of the non-strange
hadrons appropriate mesonic fields and interactions
among the baryonic and mesonic fields must be introduced in the spirit
of the Walecka model \cite{Wal77}.
This  model, the so called Quantumhadrodynamics turned out to
be the unique framework for reformulating of our whole knowledge
on nuclear physics \cite{Ser97}.
In the framework of the
 Quantumhadrodynamics
a great number of interesting
properties of the
hadronic matter has been studied rather successfully \cite{Lov94}.
Among others the role of the delta resonances has been studied very
extensively \cite{Mac87}, \cite{Wal87}.
It seems to be worthwhile to reexamine these problems in the context
of the generalised Rarita-Schwinger equations.
 We assume that the interactions among non-strange baryons are
mediated by non-strange pseudoscalar and vector mesons described
 by the isovector   $\pi_{m}(x)$ and  $\rho_{\mu m}(x)$,
and  the isoscalar  $\eta(x)$ and $\omega_{\mu}(x)$ fields.
The Lagrangian density of the system can be written as follows:
 
\begin{equation}
L=
\bar \psi (\gamma_\kappa \partial^\kappa + M_{N}
-i g_{\pi} \gamma^{5}\tau^n \pi_{ n}
-i g_{\rho} \gamma^{\kappa}\tau^n \rho_{\kappa n}
-i g_{\eta} \gamma^{5} \eta
-i g_{\omega} \gamma^{\kappa}\omega_{\kappa}) \psi \]
\[ +\bar\psi_{\mu m} (  \gamma_\kappa \partial^\kappa + M_{\Delta}
-i g_{\pi} \gamma^{5}\tau^n \pi_{ n}
-i g_{\rho} \gamma^{\kappa}\tau^n \rho_{\kappa n}
-i g_{\eta} \gamma^{5} \eta
-i g_{\omega} \gamma^{\kappa}\omega_{\kappa}) \psi_{\mu m} \]
\[ -i \bar \psi ( g_{\rho}'\rho_{\mu m} +
g_{\pi}'{\gamma^5} \partial_\mu \pi_{ m})  \psi_{\mu m}
-i \bar \psi_{\mu m}( g_{\rho}'\rho_{\mu m}+
g_{\pi}'{\gamma^5} \partial_\mu \pi_{ m})  \psi  \]
\[+L_{\pi}+L_{\rho}+L_{\eta}+L_{\omega},
\end{equation}
where $L_\pi$, $L_\rho$, $L_\eta$, and $L_\omega$ are the Lagrangians
of the free $\pi$, $\rho$, $\eta$, and $\omega$ fields, resp., and
the constraints
\begin{eqnarray}
\gamma_\mu
 \psi^{\mu m}
=0,
\\
 \tau_m
 \psi^{\mu m}
=0
\end{eqnarray}
must be taken into account.
  Here it is assumed, that
 direct
 couplings between the nucleon and
the delta
are produced by the $\rho_{\mu m}(x)$ fields and by the derivative of
the pion fields $\partial_{\mu}\pi_{m}(x)$.
 
       As it was discovered by K. Novob\'atzky
\cite{Nov38} the quantisation in the presence of constraints can be
performed in a
consistent way if the
independent degrees of freedom are separated
before the quantisation procedure
and only the independent,
so called dynamical variables are quantised. Using   this method
 the loss of the explicit covariance is the price
 we have to pay as
 a definite frame of reference must be chosen.
If we express the  field $\psi^{\mu m}(x)$ in terms of the 3/2
isospin and 3/2 helicity eigenstates
then the
constraints
are satisfied automatically,   consequently in
the
quantisation procedure only the components of the delta field are involved.

\section{Elimination of the Delta Degrees of Freedom}
\indent

     If the system is at high  temperature, as it is
the case in  high energy heavy ion collision, then the delta
contribution to the baryon density may  be substantial. At low
temperature,
however, as in the ground state of nuclear matter, the contribution of
the delta to the baryon density is much more reduced, therefore at low
temperature we may perform a formal elimination of the delta field.
 
      Starting from the coupled field equations, given by
\begin{equation}
\label{first}
(\gamma_\kappa \partial^\kappa + M_{\Delta}
-i g_{\pi} \gamma^{5}\tau^n \pi_{ n}
-i g_{\rho} \gamma^{\kappa}\tau^n \rho_{\kappa n}
-i g_{\eta} \gamma^{5} \eta
-i g_{\omega} \gamma^{\kappa}\omega_{\kappa}) \psi_{\mu m} \]
\[-i (g_{\rho}'\rho_{\mu m}+ g_{\pi}'\gamma^5 \partial_\mu \pi_{
m}))\psi=0,
\end{equation}
\begin{equation}
\label{second}
(\gamma_\kappa \partial^\kappa + M_{N}
-i g_{\pi} \gamma^{5}\tau^n \pi_{ n}
-i g_{\rho} \gamma^{\kappa}\tau^n \rho_{\kappa n}
-i g_{\eta} \gamma^{5} \eta
-i g_{\omega} \gamma^{\kappa}\omega_{\kappa}) \psi \]
\[-i(g_{\rho}'\rho_{\mu m} +
g_{\pi}' \gamma^5 \partial_\mu \pi_{ m})  \psi_{\mu m} =0,
\end{equation}
we can solve  Eq. (\ref{first}) for  $\psi_{\mu m}$:
\begin{equation}
\psi_{\mu m}=
(   \gamma_\kappa \partial^\kappa + M_{N}
-i g_{\pi} \gamma^{5}\tau^n \pi_{ n}
-i g_{\rho} \gamma^{\kappa}\tau^n \rho_{\kappa n}
-i g_{\eta} \gamma^{5} \eta
-i g_{\omega} \gamma^{\kappa}\omega_{\kappa})^{-1}  \]
\[ i(g_{\rho}'\rho_{\mu m}+ g_{\pi}' \gamma^5 \partial_\mu \pi_{
m}))\psi
\end{equation}
and  substituting this into
 Eq. (\ref{second}), we obtain:
\begin{equation}
(   \gamma_\kappa \partial^\kappa + M_{N}
-i g_{\pi} \gamma^{5}\tau^n \pi_{ n}
-i g_{\rho} \gamma^{\kappa}\tau^n \rho_{\kappa n}
-i g_{\eta} \gamma^{5} \eta
-i g_{\omega} \gamma^{\kappa}\omega_{\kappa}+{\Phi})
\psi =0,
\end{equation}
where the term $ \Phi$ is defined as
\begin{equation}
\Phi=
 (g_{\rho}'\rho_{\mu m}+ g_{\pi}' \gamma^5\partial_\mu \pi_{ m}) \]
\[ (\gamma_\kappa \partial^\kappa + M_{\Delta}
-i g_{\pi} \gamma^{5}\tau^n \pi_{ n}
-i g_{\rho} \gamma^{\kappa}\tau^n \rho_{\kappa n}
-i g_{\eta} \gamma^{5} \eta
-i g_{\omega} \gamma^{\kappa}\omega_{\kappa})^{-1} \]
\[ (g_{\rho}'\rho_{\mu m}+ g_{\pi}' \gamma^5 \partial_\mu \pi_{ m}) .
\end{equation}
Using a complete set of 8x8 matrices the term $\Phi$ can be expanded.
 This decomposition 
produces contributions to the interaction terms of Eq. (65). The contribution 
to $M_N$ can be identified with the sigma field introduced by Walecka
 \cite{Wal77}.
As it is seen  the elimination of the degrees of freedom associated
with the delta gives rise to an effective scalar-isoscalar
interaction.
      The procedure followed here seems
to be a natural way for the introduction of the `nonexisting' sigma
field,
furthermore it supports the traditional interpretation of the sigma
particle as a correlated meson pair.

\vskip 10pt
\noindent \large {\bf Acknowledgement}\normalsize
\vskip 10pt
\noindent
This work has been supported by the Hungarian Academy of Sciences,
the Deutsche Forschungs Gemeinschaft and the Hungarian Research Fund OTKA.
\\
\[   \] \\
E-mail: lovas@heavy-ion.atomki.hu\\
\[   \]     \\

\vfill\eject
\end{document}